\documentclass{iau}
\usepackage{graphicx}

\title[Void Dynamics] 
{Void Dynamics}

\author[Padilla et al.]   
{Nelson D. Padilla$^{1,2}$,
Dante Paz$^3$,
Marcelo Lares$^3$,
Laura Ceccarelli$^3$,
Diego Garc\'\i a Lambas$^3$,
Yan-Chuan Cai$^4$
 \and Baojiu Li$^4$}

\affiliation{$^1$Instituto de Astrof\'\i sica, \\ Universidad Cat\'olica de Chile, \\
Vicu\~na Mackenna 4860, Santiago, Chile \\ email: {\tt npadilla@astro.puc.cl} \\[\affilskip]
$^2$Centro de Astro-Ingenier\'\i a, \\ Universidad Cat\'olica de Chile, \\
$^3$Instituto de Astronom\'\i a Te\'orica y Experimental (IATE), \\Laprida 922, C\'ordoba, Argentina\\
$^4$Institute for Computational Cosmology\\Durham University\\South Road, Durham, DH1 3LE, UK\\
}

\pubyear{2008}
\volume{308}  
\pagerange{119--126}
\jname{The Zeldovich Universe}
\editors{R. van de Weygaert, S. Shandarin, e. Saar \& J. Einasto, eds.}
\begin{document}

\maketitle

\begin{abstract}
Cosmic voids are becoming key players in testing the physics of our Universe.
Here we concentrate on the abundances and the dynamics of voids as these are among the best candidates 
to provide information on cosmological parameters.  Cai, Padilla \& Li (2014)
use the abundance of voids to tell apart Hu \& Sawicki $f(R)$ models from General Relativity.  An interesting
result is that even though, as expected, voids in the dark matter field are emptier in $f(R)$ gravity due to the fifth force expelling 
away from the void centres, this result is reversed when haloes are used to find voids.  The abundance of voids in this case
becomes even lower in $f(R)$ compared to GR for large voids.  Still, the differences are significant and this
provides a way to tell apart these models.  The velocity field differences between $f(R)$ and GR, on the other hand, are 
the same for halo voids and for dark matter voids.
Paz et al. (2013), concentrate on the velocity profiles around voids.  First they show the necessity
of four parameters to describe the density profiles around voids given two distinct void
populations, voids-in-voids and voids-in-clouds.  This profile is used to predict peculiar velocities around voids,
and the combination of the latter with void density profiles allows the construction of model
void-galaxy cross-correlation functions with redshift space distortions.  When these models
are tuned to fit the measured correlation functions for voids and galaxies in the Sloan
Digital Sky Survey, small voids are found to be of the void-in-cloud type, whereas larger
ones are consistent with being void-in-void.  This is a novel result that is obtained
directly from redshift space data around voids.  These profiles can be used to
remove systematics on void-galaxy Alcock-Pacinsky tests coming from redshift-space distortions.
\end{abstract}

\firstsection 
\section{Introduction}

Cosmic voids are underdense regions in the universe which occupy a significant fraction of the total volume, and as such
are potentially powerful tools for statistical studies of the Universe.  They are the result
of the history of growth of perturbations since, the faster the virialised structures gain mass, the faster
other parts of the Universe must empty of material to feed this mass increase.  Therefore, statistics
of the abundance and the rate of growth of voids are related to the growth factor in the universe,
which in turn is related to cosmology.  Regarding the rate of growth of voids, it is possible
to study it via the evolution of the abundance of voids, or directly measuring velocity profiles
around voids.

The abundance of voids has been proposed as a way to test cosmology.  However, this has proved to be challenging
given the difference in abundance obtained using different void finders (e.g. Colberg et al. 2005).  
Jennings et al. (2013) have recently obtained a $\sim 16$ percent agreement between simulations
and excursion set predictions (see for instance Sheth \& van de Weygaert, 2004), but 
in their case they use the dark matter particles to do this comparison.

The dynamics around voids has been studied in simulations and data (e.g., Padilla et al. 2005,
Ceccarelli et al., 2006, respectively) where the amplitude of peculiar velocities
was shown to reach a maximum of a few hundreds of $kms^{-1}$, and has now been included in studies of density profiles
of voids in the directions parallel and perpendicular to the line of sight; if the voids are spherical and
the effects from peculiar velocities are either consistently similar for different types of voids, or can
be modeled, these can be used to disaffect the redshift space distortions on the profile along the line
of sight.  Once this is done, the remaining differences in the two directions can be used to study
cosmology using the Alcock-Pacinsky test (e.g. Lavaux \& Wandelt 2012, Sutter et al. 2014).  

However, there is still debate on whether the profiles (density and velocity) of voids are self-similar, of if
they separate into (at least) two different families of voids.  Sheth \& van de Weygaert (2004) propose the existence of two distinct types of voids.  The ones called void-in-void which are underdense up to
very large distances from the void centre (several void radii), and those that are embedded in a larger overdensity,
called void-in-cloud.  It is clear that the density profiles of these two types will be different.
Ceccarelli et al. (2013) selected
voids that lie in an overdensity at tree times the void radius as voids-in-clouds, and voids that do not lie
in an overdensity at this same radius as voids-in-voids and showed that
the velocity profiles of these two families of voids are different.  Void-in-clouds show an infall of matter
beyond three times the void radius.  Void-in-voids only show outflows.  Therefore, in order to be able to
do high precision Alcock-Pacinsky tests it will be necessary to take into account these differences fully.

In this proceedings we will first briefly discuss the void finding algorithm adopted for the
works whose results will be presented here; then we will show very recent results on the abundance of voids near the present epoch of
the universe for the $\Lambda$CDM universe but also for alternative gravity $f(R)$ models.  We will also
show the first measurements of void expansion using real data by means of fitting the
void-galaxy cross-correlation function in the spectroscopic Sloan Digital Sky Survey (SDSS).  The works
we concentrate on are Cai, Padilla \& Li (2014) and Paz et al. (2013), respectively.

\section{Finding voids}

The works we refer to in this proceeding use voids identified using modified versions of the Padilla et al. (2005, P05 from now on)
finder.  This finder is explained in detail in P05, so we only include a brief description of the algorithm.  The steps
that are followed are first, to take a number of prospective void centres (either by making a full spatial search or starting
from low density regions), and grow spheres around these until the density within them reaches a threshold value.  For Paz et al.
(2013, P13 from this point on)
and Cai, Padilla \& Li (2014, C14) these thresholds were $0.1$ and $0.2$ times the average density of the mass or tracers, respectively.  The size
of the sphere at this point is defined as the void radius.  

Depending on whether the tracers are sparse or not, a minimum number of tracers (DM particles or galaxies in C14 and P13, respectively)
is required for a void to be accepted into the final sample.  In the case of using massive haloes in C14, no minimum number of 
haloes is required to lie within the voids.

The modifications to the P05 finder in each case, C14 and P13, are described in detail in each paper.  In C14 the modifications improve the
convergence with numerical resolution and is adapted to be applied to massive haloes, which suffer from being sparse
samplers of the density field.  In particular, since no minimum number of haloes within voids is required 
when haloes are used to detect voids, a new parameter, $\sigma_4$ is defined,
$$\sigma_4\equiv\sqrt{\left< \left(\frac{d_i-\left< d_i\right>_4}{\left< d_i\right>_4}\right)^2 \right>_4},$$
that measures the dispersion of the distance between the void centre and the nearest four haloes.  The smaller this parameter
is the better defined the void centre is, and the more spherical its shape is.
In P13 the modifications concentrate on adapting the finder to work in redshift space
in a survey affected by a complicated angular mask (though it is applied to volume limited samples).  The space density
of galaxies used by P13 makes the sparseness not as important as in C14.

\begin{figure}[b]
\begin{center}
 \includegraphics[width=2.8in]{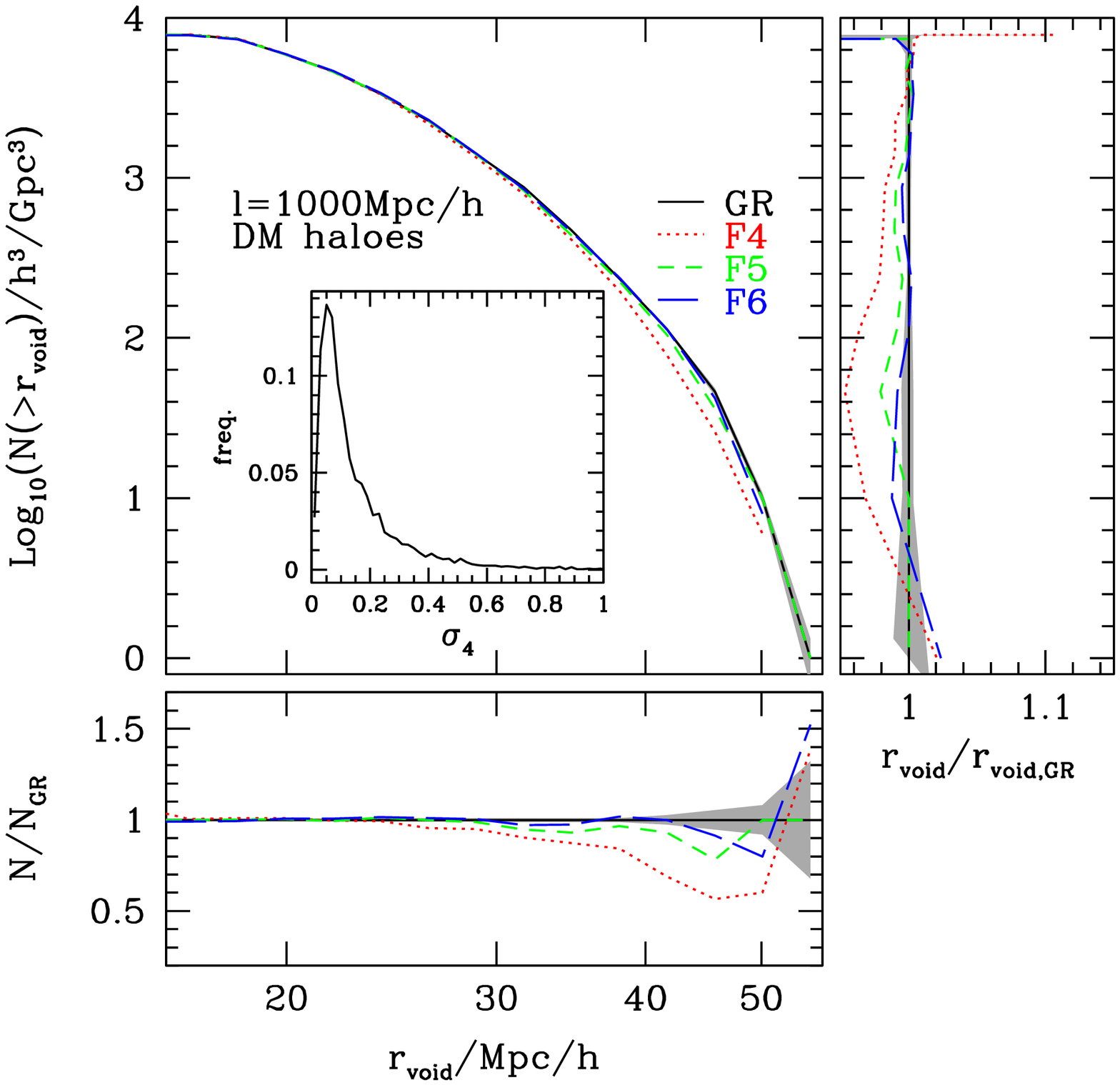} 
\hskip -.5cm
\includegraphics[width=2.6in]{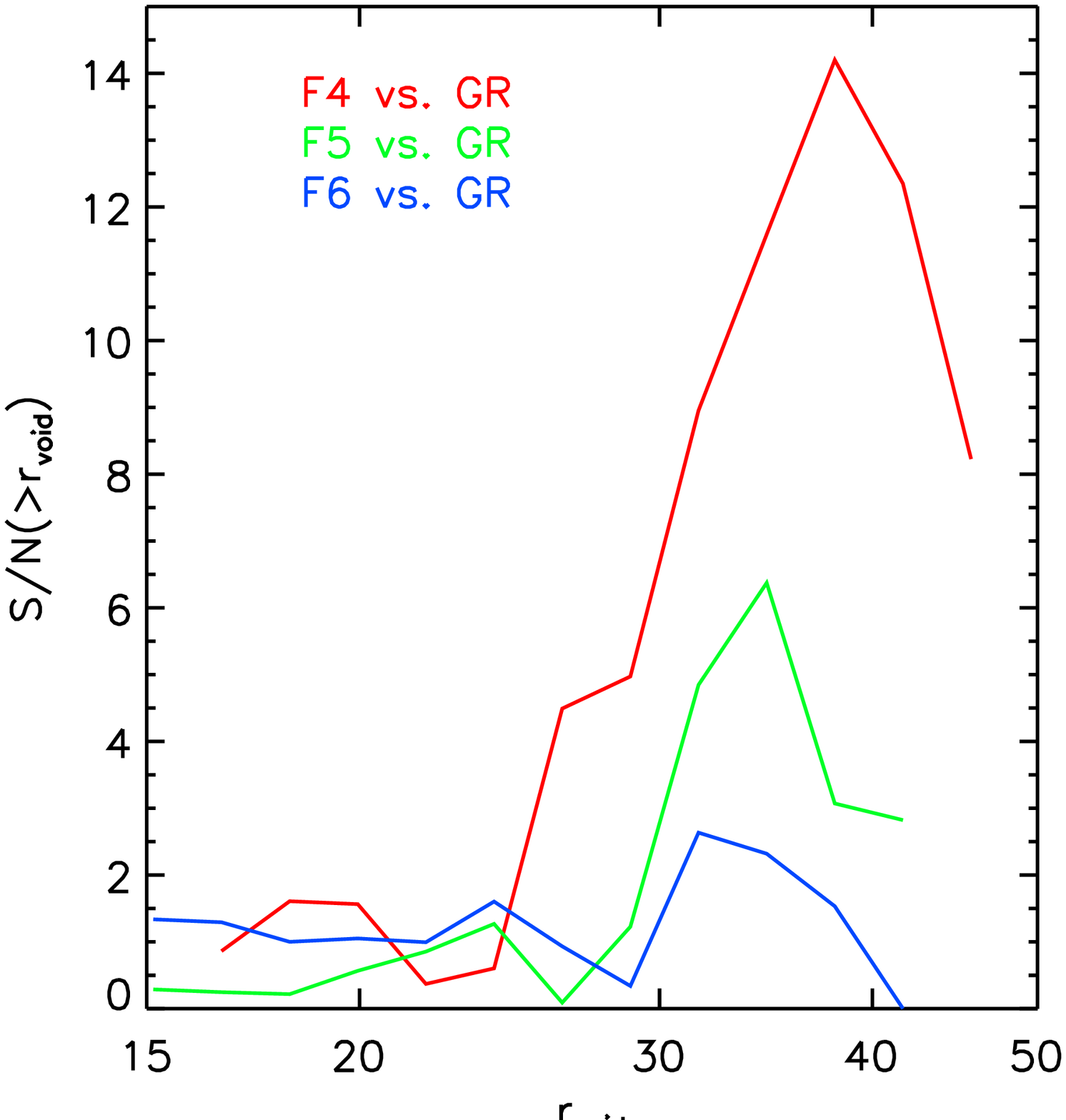}
 \caption{
Left: Void abundances when dark matter haloes are used as tracers of the density
field.  The main panels shows the abundances of voids above a minimum void radius, with
differently line types and colours for different gravity models as indicated in the figure key.
The right panel shows the ratio of abundances at fixed cumulative density.  The bottom panel
shows the ratio of abundances at a fixed void radius.
The inset in the main
panel shows the distribution function of $\sigma_4$, the relative dispersion in the distance
to the first four haloes above the void density threshold, for GR only; a small value of $\sigma_4$ indicates
a well centred void.  
Right: The corresponding S/N from void abundances.
}
   \label{fig:dndrhaloes}
\end{center}
\end{figure}

\section{Void abundance}

If the interest is to compare model predictions with different cosmologies or different recipes
for gravity, it is valid to use a single void finding algorithm.  This is the spirit with which C14 search
for ways to tell apart modified $f(R)$ gravity models from General Relativity.  By using the P05 with
the modifications we mentioned earlier, they are able to find voids in simulations with GR and
with Hu \& Sawicki (2007) $f(R)$ models with different amplitudes of the scalar field strength
of $|f_{R0}|=10^{-6}$, $10^{-5}$ and $10^{-4}$ (referred to as F6, F5 and F4 from this point on).  The simulations cover a volume of $1$ cubic
Gigaparsec, and all share the same initial conditions, and were run with the ECOSMOG code (Li et al., 2012).  
This way all the differences found in the resulting voids arise solely from the different recipes for gravity.

C14 first identified voids in the matter density field traced by dark matter particles, and they were
able to confirm analytic expectations that in $f(R)$ models, voids tend to be emptier due to the
fifth force acting in the opposite direction to that of Newtonian gravity (Clampitt et al. 2013).
The differences in abundance between F4 and GR are above $100$ percent with extremely high significances.

However, when using haloes with masses $\gtrsim 10^{13}$h$^{-1}M_{\odot}$ (the lower limit in mass
is slightly different in each model so that the number density of haloes is the same), they found the result
shown in the left panels of Figure \ref{fig:dndrhaloes}.  In the figure it is clear that the voids found in $f(R)$ (large ones in particular)
are about as abundant as in GR.  C14 interpret this result as a subtle effect of the fifth force
that could only be fully realised with the non-linear evolution of the density field as it evolved
in the numerical simulation.  The fifth force, locally, induces the formation of haloes.  The structures
embedded in voids are therefore located in a region where this force is increased, which causes more
massive haloes to form within voids in $f(R)$ in comparison to GR.  This makes voids even smaller
than in GR when haloes are used to identify them.  

The small differences are still significant, to the levels shown in the right main panel in Figure \ref{fig:dndrhaloes}, where
the peak in signal-to-noise ratio, $S/N$, is found for haloes of around $30-40$h~$^{-1}$Mpc, and is
$S/N\sim 2$ even for F6. Therefore, the abundance of voids can be used to tell apart very
mild fifth force models such as F6, although with a rather small significance.  We refer the reader to C14 for
a full discussion of the implications of this result, as well as for conclusions regarding the
density and velocity profiles of voids in GR and $f(R)$ models.

\section{Void dynamics}

\begin{figure}
\center{ \includegraphics[width=.8\textwidth]{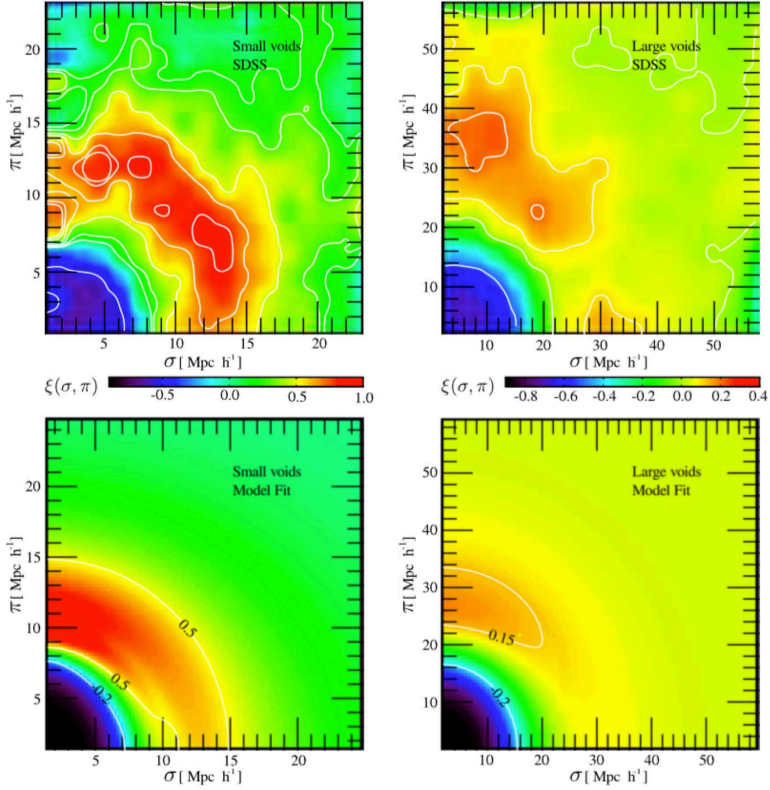}}
\vskip -1.7cm
\caption{
   Redshift space distortions of galaxies in the SDSS (upper left panel) and
   the best fit models (bottom left panels) for small voids-in-clouds  
   ($6<\mathrm{R}_{\mathrm{void}}/$Mpc$h^{-1}<8$).  Larger voids
   ($10<\mathrm{R}_{\mathrm{void}}/$Mpc$h^{-1}<20$) are shown correspondingly on the
right panels.
}
\label{F_SDSS_xisigpi} 
\end{figure}

\begin{figure}
\center{\includegraphics[width=.8\textwidth]{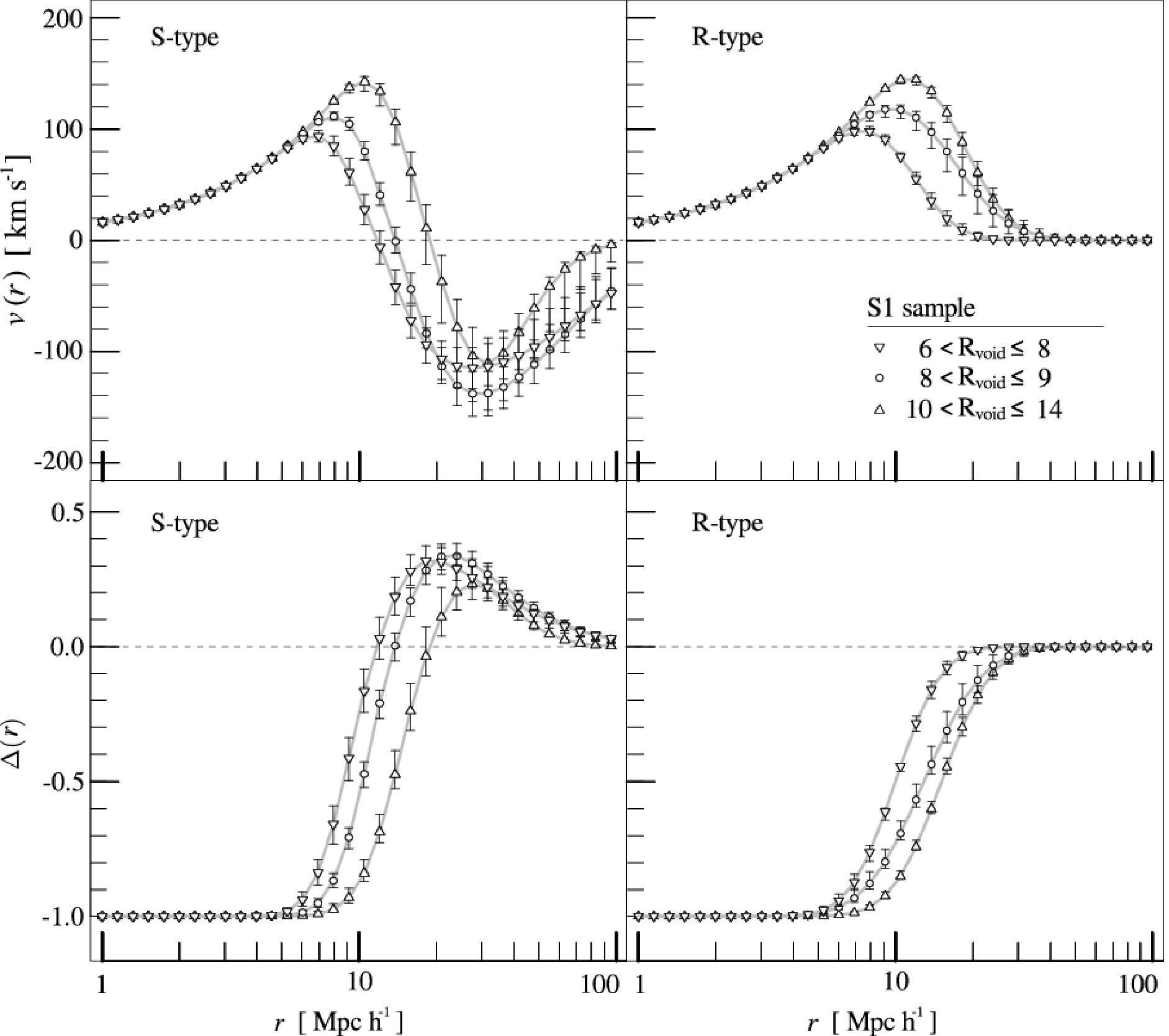}}
\caption{
Void--centric radial galaxy density profiles $\Delta(r)$ (lower
panels) and void--centric radial galaxy velocity profiles (upper panels) 
for voids-in-clouds (left) and voids-in-voids (right).
Different void radii ranges are indicated with different symbols as shown in the key.
Error bars indicate the region enclosing all curves within 68.3\%
uncertainty in the parameter space.  The sample of voids in this case
corresponds to sample S1 in P13 which is volume limited out to a redshift $z=0.08$.
}
\label{F_SDSS_profiles} 
\end{figure}

Once the abundance of voids is determined, one direct application that can be followed is to study the
dynamics of voids.  As shown by C14, when moving from the density field of the mass, to that
traced by haloes, the differences between $f(R)$ and GR become smaller.  This, however, does not
happen when studying velocity profiles.

In a previous paper, P13 studied the dynamics in detail for GR voids.  We will concentrate on these
results in what follows.

Before going to the actual dynamics, it should be noticed that the source of peculiar velocities
is the distribution of mass.  Therefore, it is necessary to briefly discuss some
aspects of density profiles around voids.
Several recent works (e.g. Nadathur et al., 2014) argue that the density profiles around voids
are self similar.  If this is the case, then one can adopt self-similar density profiles
to model the void velocities.  However, Sheth \& van de Weygaert (2004) showed that
there are probably voids of the void-in-cloud and void-in-void types.  
Ceccarelli et al. (2013, C13) adopted a practical definition to separate voids into these
two classes.  If the integrated overdensity in spheres of $3$ times the void radius
is positive, then the void is of the void-in-cloud family.  If it is negative, it is
a void-in-void.  Using simulations C13 then showed that indeed, profiles are different for these two
families.  However, C13 also find that the largest voids, those with radius $>10$h$^{-1}$Mpc,
are mostly of the void-in-void type.  Therefore, for sparse samples where voids are
rather large, the approximation of a self-similar profile is adequate.
P13 study the dynamics for voids of sizes going down to
$6$h$^{-1}$Mpc and the distinction between void-in-cloud and void-in-void
is unavoidable.

In order to model the integrated density profiles of voids, P13
introduce a simple empirical model that contains the necessary
features of the density profiles of the two types.
The voids-in-voids have the
simplest profile shapes, a continuously rising function to the mean density.
The error function $\mathrm{erf}(x)$ behaves as needed,
\begin{equation}
	\label{eq_rho_r}
	\Delta_R(r) =
   \frac{1}{2}\left[\mathrm{erf}\left(\mathrm{S}\;\mathrm{log}(r/\mathrm{R})\right)-1\right].
\end{equation}
This model depends on two parameters, the void radius R and a
\textit{steepness} coefficient S.
On the other hand, the profiles of  voids-in-clouds are more
complex and require two additional parameters in order to account
for the overdensity shell surrounding the voids.
To the rising term in Eq. \ref{eq_rho_r}, P13 add a
peak in the density profile, 
\begin{equation}
	\label{eq_rho_s}
	\Delta_S(r) =
   \frac{1}{2}\left[\mathrm{erf}\left(\mathrm{S}\;\mathrm{log}(r/\mathrm{R})\right)-1\right] + 
   \mathrm{P}\,\mathrm{exp}\left(-\frac{\mathrm{log^2}(r/\mathrm{R})}{2\Theta^2(r)}\right)
\end{equation}
where the Gaussian peak has an asymmetric width, 
\begin{equation}
   \Theta(r) = \left\{
   \begin{array}{ll}
      1/\sqrt{2\;\mathrm{S}} & r<\mathrm{R}\\
      1/\sqrt{2\;\mathrm{W}} & r>\mathrm{R}
   \end{array}
\right.\nonumber
	\label{eq_rho_Theta}
\end{equation}
This allows them to modify the size of the shell through the W parameter,
without changing the inner shape of the profile, related to the parameter S. 
In this way, a cumulative profile for a void-in-cloud requires four 
parameters, R, S, P and W.

P13 use these profiles to obtain the average radial velocity profiles around voids using
the formalism by Peebles (1979).  To fit the measured cross-correlations it is then necessary to convert the 
density profile into
a correlation function.  This correlation function is then calculated as a function
of two separations, one in the direction of the line of sight, and in the one perpendicular to it.
Before adding any velocity information, the correlation function is identical in these two directions.
Then the correlation function in the direction of the line of sight
 is convolved with a peculiar velocity model that responds to the density of matter out
to a given separation, and with a gaussian distribution of velocities to take into account
random motions within groups of galaxies in overdensities around the voids.   For the full
details of the calculations we refer the reader to P13.

P13 apply their method to volume limited samples of SDSS, where they identify voids using the modified
P05 finder (modifications explained in detail in C13), and measure the void-galaxy cross-correlation
functions for the resulting samples.  The galaxies used in this measurement are obtained from
the full flux limited sample.  The results for two sets of voids are shown in the top panels of Figure \ref{F_SDSS_xisigpi}.
As can be seen, the correlation function is negative close to the void centres and
at the void radius it quickly rises to reach positive values (red regions).  These regions are slightly elongated
along the line of sight.  This shows that there is indeed a velocity dispersion in
the void walls, otherwise this ridge would look just as wide in the two directions (parallel and perpendicular to the line
of sight).  Also, larger voids show a less pronounced ridge (right panels) as these are
mostly of the void-in-void type.

P13 study samples of small and large voids which
effectively separates the problem
into void-in-cloud and void-in-void centres, respectively; note that as the real space profiles are unknown for redshift space
data, this is only an expectation that can be confirmed (or not) by looking at the resulting profiles after fitting the
observed correlations.  The bottom left panel show the result of fitting
for the profile parameters, such that the model cross-correlation functions match as closely as possible
the measured ones. As can be seen, the model cross-correlation function captures most of the phenomenology
seen in the observations.

The profiles corresponding to the best fit parameters are shown in Figure \ref{F_SDSS_profiles}.  As can be seen
the measured cross-correlations are clearly consistent with a picture where small voids have an overdense ridge, and
a velocity profile that shows outflow out to the void radius, and then infall at larger separations.  This is a clear
signature of a void-in-cloud centre.  On the other hand, larger voids point to profiles consistent with
that of void-in-void centres, with a smoothly rising profile, and outflow velocities at all radii; we show the
results of the fit for larger voids in the right panels.  This constitutes
the first detection of the two types of voids in actual data, although in an indirect way that uses a linear theory model
for the velocity profiles.  This, however, implies that in order to use void-galaxy correlations to perform
Alcock-Pacinsky type tests, one needs to be careful to take into account the proper peculiar velocity field
around voids to remove systematics from redshift-space distortions.

\section{Conclusions and outlook}

Voids are promising structures with a great potential to provide important constraints on cosmology and gravity.
Even though there is no clear convergence on the abundance of voids (e.g. Colberg et al., 2005) and
therefore this statistics is not suitable for cosmological constraints yet, there are other applications
that show excellent capabilities in this sense.

In Cai, Padilla \& Li (2014), a single void finder is used, that of Padilla et al. (2005) with modifications
to allow it to run in parallel and also to provide stable results for sparse samples of haloes.  They use
this finder to search for voids in simulations with different gravity recipes, corresponding to General
Relativity, and to three Hu \& Sawicki $f(R)$ models with different strengths of the scalar field
parameter $|f_{R0}|=10^{-6}$, $10^{-5}$ and $10^{-4}$ (F6, F5 and F4).  All the simulations follow a volume
of $1$ cubic Gigaparsec to $z=0$ and start from the same initial conditions.  The differences between 
the populations of voids respond solely to the different strengths of the fifth force (absent in GR).
They find that,  when the dark matter field (particles) are used
to detect voids, those in $f(R)$ models are larger as expected (Clampitt et al., 2013).  However, when 
using massive haloes of $\gtrsim 10^{13}$h$^{-1}M_{\odot}$
as tracers, a surprising result arises.  Voids are just as abundant in $f(R)$; furthermore, when restricting
the comparison to only the largest voids with radii $>30-40$h~$^{-1}$Mpc, $f(R)$ voids are less abundant.
This difference, although not as large as for voids found from the dark matter field, is still enough
for a significant detection of the effect of the fifth force in F5 and F4, and marginally for F6,
for a $1$ cubic Gigaparsec volume (the significance would be larger for volumes such as those of
LSST or EUCLID).  

It is important to understand why the abundance of voids found using haloes is lower in $f(R)$.  The
analysis by C14 indicates that this is due to the fact that the fifth force is stronger in voids (low
densities) and this increases the growth of haloes in these regions.  Therefore, haloes in voids tend
to be more massive in $f(R)$.

In order to tell appart $f(R)$ from GR, a single finder can be used, but it would
need to be tested on detailed mock catalogues with the different gravity recipes.  Then, it would
be possible to search for the gravity that
best matches the observed abundances which would be found using the same void finder in actual data.

C14 also point out that the different results found when using the dark matter field compared to
using tracer haloes are not present when studying velocity profiles.  

To obtain velocity profiles, though, it is of great importance to determine the overdensities surrounding
the voids, as these are the source of peculiar velocities.  In Paz et al. (2013) they set out 
to use redshift space distortions around voids to learn about the density profiles around voids
of the void-in-cloud and void-in-void types (Sheth \& van de Weygaert, 2004).  P13 
propose a four parameter function that describes the full range of void profiles, for these
two void types.
They use this functional form to obtain redshift space distortions around voids, by calculating
the cross-correlation function between voids and galaxies as a function of the separations
perpendicular and parallel to the line of sight.  When they fit these parameters to obtain
models that best reproduce the measured cross-correlations from voids and galaxies in the Sloan
Digital Sky Survey, they find that small voids are best fit by void-in-cloud type profiles, whereas
large voids are consistent with the void-in-void type.  This represents the first indirect detection
of these two types of voids using redshift space data.

An interesting application of these predictions for redshift space distortions around voids
is to use them to perform Alcock-Pacinsky effect tests (Lavaux \& Wandelt 2012, Sutter et al. 2014),
but removing the distortions coming from peculiar velocities, which could otherwise systematically change
the shapes of the voids in the direction parallel to the line of sight.

\end{document}